\documentclass[aps,superscriptaddress,amssymb,nobibnotes]{revtex4-2}
\usepackage{amsmath,bm}
\usepackage{amssymb}
\usepackage{graphicx}
\usepackage{tabularx}
\usepackage{url}
\usepackage[usenames,dvipsnames]{xcolor}
\usepackage{hyperref}
\hypersetup{colorlinks=true, linkcolor=BrickRed, urlcolor=blue!50!black, citecolor=blue!50!black}
\usepackage[capitalize]{cleveref}
\usepackage{cleveref}
\AtBeginDocument{}
\crefrangelabelformat{equation}{(#3#1#4)$-$(#5#2#6)}
\begin{document}

\title{
Aging during Phase Separation in Long-Range Ising Model
}

\author{Soumik Ghosh}
\author{Subir K. Das}
\email{das@jncasr.ac.in}
\affiliation{Theoretical Sciences Unit and School of Advanced Materials, Jawaharlal Nehru Centre for Advanced 
Scientific Research, Jakkur P.O., Bangalore 560064, India}
\date{\today}


\begin{abstract}
The kinetics of domain growth and aging in conserved order parameter systems, in the presence of short-range interaction, is widely studied. Due to technical difficulties and lack of resources, regarding computation, the dynamics is still not well established in the cases where long-range interactions are involved.\hspace{0.1cm}Here we present related results from the Monte Carlo simulations of the two-dimensional long-range Ising model (LRIM). Random initial configurations, for $50:50$ compositions of up and down spins, mimicking high temperature equilibrium states, have been quenched to temperatures inside the coexistence curve. Our analysis of the simulation data, for such a protocol, shows interesting dependence of the aging exponent, $\lambda$, on $\sigma$, the parameter, within the Hamiltonian, that controls the range of interaction. To complement these results, we also discuss simulation outcomes for the growth exponent. The obtained values of $\lambda$ are compared with a well-known result for the lower bounds.  For this purpose we have extracted interesting properties of the evolving structure.
\end{abstract}


\maketitle

\section{Introduction}
Typically, following a perturbation, relaxation in an older system occurs slower than in a younger system. Such aging phenomena, in an evolving system, during a phase transition, is often studied via the autocorrelation function \cite{Fisher1988, Liu1991, Yeung1996, Henkel2004, Lorenz2007, Zanetti2009, Midya2014, Midya2015, Christiansen2020}
\begin{equation}
    C_{ag}(t,t_w)=\langle\psi(\vec{r},t)\psi(\vec{r},t_w)\rangle-
    \langle\psi(\vec{r},t)\rangle\langle\psi(\vec{r},t_w)\rangle.
\label{Cag_eq}
\end{equation}
Here $\psi$ is a space ($\Vec{r}$) and time-dependent order parameter, the symbols $t$ and $t_w$ representing, respectively, the observation and the waiting times. The latter is also called the age of the system. The slower decay of $C_{ag}$ for an older system is a violation of the time-translation invariance. This implies that in an ``away from steady-state" situation there is no scaling or collapse of data for $C_{ag}$ when results for different $t_w$ are plotted versus $t-t_w$. However, collapse is observed \cite{Fisher1988} when the data are plotted versus $t/t_w$. In the limit $t/t_w\rightarrow\infty$, one then discusses the scaling behaviour \cite{Fisher1988} 
\begin{equation}
    C_{ag}(t,t_w) \sim \left( \dfrac{t}{t_w}\right)^{-\lambda^t}.
    \label{alpha_lambda}
\end{equation}

Note that during a phase transition, as a homogeneous or disordered system is quenched to a miscibility gap or ordered region of the phase diagram, domains, rich or poor in specific components, form and their average size, $\ell$, grows as \cite{Bray2002}
\begin{equation}
    \ell \sim t^\alpha.
    \label{alphs}
\end{equation}
One further defines $\lambda=\lambda^t/\alpha$, such that $C_{ag}(t,t_w)\sim (\ell/\ell_w)^{-\lambda}$, where $\ell_w$ is the value of $\ell$ at $t=t_w$. The quantities $\alpha$ and $\lambda$ are two important power-law exponents in the literature of kinetics of phase  transitions \cite{Fisher1988,Zanetti2009,Bray2002,Nalina2019}. These are referred to as the growth and the aging exponents, respectively. For many systems, with short-range interactions \cite{Fisher1988, Liu1991, Yeung1996, Henkel2004, Lorenz2007, Zanetti2009, Midya2014, Midya2015, Christiansen2020, Das2017, Saikat2020, Nalina2019, Lowen2017, Amar1988}, particularly for the Ising model (IM) \cite{Fisher1988, Yeung1996, Lorenz2007, Zanetti2009, Midya2014, Midya2015, Nalina2019, Amar1988}, these have been estimated fairly accurately. Despite the IM being one of the most popular prototype models for studies of phase transition, progress is very limited for the long-range (LRIM) variety.

The Ising model Hamiltonian can be written in the general form \cite{Zanetti2009,Christiansen2020,Bray2002}
\begin{equation}
    H=-\frac{1}{2}\Sigma_i \Sigma_{j\ne i} J_{ij}S_iS_j ,
\label{hamiltonian}
\end{equation}
where $J_{ij}$ is the strength of interaction between spins $S_i$ and $S_j$ ($=\pm 1$), sitting at the lattice points $i$ and $j$. For $J_{ij}> 0$, one expects mostly parallel alignment of spins, at low enough temperatures. For standard purposes \cite{Bray2002}, one considers $J_{ij}=J$ and terminates the interaction at the nearest neighbour distance.\hspace{0.1cm}For defining LRIM, on the other hand, a power-law variation of the strength, as a function of $r$, the inter-site distance, is considered \cite{Bray1993,Corberi2019}:
\begin{equation}
    J_{ij}=\dfrac{J}{r^{d+\sigma}},
    \label{J_form}
\end{equation}
where $d$ is the spatial distance and $\sigma$ is a constant. In the equilibrium context, the value of $\sigma$, that separates the short-range and long-range universality classes, is close to two \cite{Fisher1972}. For problems associated with kinetics, such a boundary is set \cite{Bray1993} at $\sigma=1$.

For nonconserved order-parameter dynamics, that is relevant for ordering, say, in magnetic systems, it is predicted \cite{Bray1993} that 
\begin{equation}
    \alpha =
                    \begin{cases}
                        \frac{1}{1+\sigma},  & \mbox{for}\  \sigma < 1 \\
                        
                        \frac{1}{2},           & \mbox{when}\ \sigma>1.
                        
                    \end{cases}
\label{alpha_noncon}
\end{equation}
Similar predictions exist for kinetics of phase separation, say, in a binary (A + B) mixture, with conserved order-parameter, as well \cite{Bray1993, Bray1994}:
\begin{equation}
    \alpha =
                    \begin{cases}
                        \frac{1}{2+\sigma},  & \mbox{for}\  \sigma < 1 \\
                        
                        \frac{1}{3},           & \mbox{when}\ \sigma>1.
                        
                    \end{cases}
\label{alpha_con}
\end{equation}
There exist certain logarithmic behaviour \cite{Bray1993} for $\sigma=1$. There are no such general predictions for $\lambda$. In fact, for $\lambda$, existing theoretical predictions are only for the nearest neighbour case with non-conserved dynamics \cite{Fisher1988,Liu1991}. For this quantity, not only the theoretical calculations, but also the computer simulations, and their analysis, are challenging, irrespective of the range of interaction, particularly for the conserved dynamics. 

For the long-range interaction, there has been a surge of interest in recent times \cite{Christiansen2020,Christiansen2021,Muller2022,Agrawal2022,Corberi2019}. However, the focus has been on the nonconserved variety. Only two studies \cite{Muller2022,Ishibara1994}, including a recent one \cite{Muller2022}, to the best of our knowledge, investigated the conserved case, reporting results on domain growth. In this work, we present results on the \textit{aging phenomena} for the conserved variety, for a wide range of values of $\sigma$, in $d=2$. To analyse the results on aging, it becomes necessary to calculate $\ell$. Our results on this latter quantity are in good agreement with the above mentioned recent report \cite{Muller2022}. We also discuss the results on $\lambda$ against a well-known bound \cite{Yeung1996}. 

\section{Model and Method}
We have considered spins on periodic square lattices of size $L\times L$. Unless mentioned otherwise, the results are presented for $\L=256$, the unit being the lattice constant. Conserved order-parameter dynamics requires that the total number of $+1$ (or particles of type $A$) and $-1$ (or particles of type $B$) spins remain constant throughout the evolution of the system. To ensure this, Kawasaki spin-exchange dynamics is used in our Monte Carlo simulations \cite{Landau2005}. In this process, two (nearest) neighbouring sites are randomly chosen and the corresponding spin states are exchanged with a certain probability following the Metropolis criterion \cite{Landau2005}. For this purpose, energy change is needed to be calculated. With the Hamiltonian in Eq. (\ref{hamiltonian}), and the coupling term mentioned in Eq. (\ref{J_form}), this calculation is computationally demanding \cite{Christiansen2019}. To minimize the cost of computation, a generalized \cite{Horita2017} Ewald summation \cite{Allenbook, Horita2017} technique is used. We have carried out the calculations with our (in-house)  \textit{parallel} codes, written with OpenMP and MPI, for even faster computation. Time in our simulations is measured in units of Monte Carlo steps (MCS), one MCS being equivalent to $L^2$ trials. For each $\sigma$ value, random initial configurations, with equal concentrations of up and down spins, are quenched to a temperature that is $0.6$ times the corresponding critical temperature \cite{Horita2017}. At a finite temperature there exists noise in the structure. This was removed, via a majority spin rule \cite{Nalina2019}, for the calculation of lengths. The latter quantity was obtained from the domain-size distribution function. All results are presented after averaging over $80$ initial configurations. For each value of $\sigma$, total run length was $t=8\times 10^4$ MCS.

\section{Results}
\begin{figure}[h]
\centering
\includegraphics*[width=0.48\textwidth]{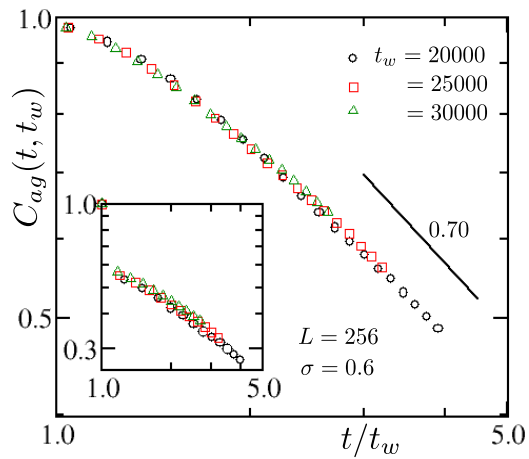}
\caption{Plots of the autocorrelation function, $C_{ag}(t,t_w)$, as a function of $t/t_w$, for LRIM with $\sigma=0.6$. Results for a few different ages have been included. The inset contains the actual data. Results in the main frame are scaled by a pre-factor, after discarding the jump,  such that $C_{ag}$ smoothly approaches $1$ as $t/t_w\rightarrow 1$. The solid line represents a power-law decay with the mentioned value of the exponent. 
}
\label{Cag}
\end{figure}

In Fig. \ref{Cag} we show a few plots of the autocorrelation function, with the variation of $t/t_w$, for $\sigma=0.6$. Original results are presented in the inset.\hspace{0.1cm}The jumps there have connection with equilibration of domain magnetization \cite{Zanetti2009}. In the main frame the results are presented by discarding this feature, keeping data having connection only with the growth of domain. Essentially, we have rescaled the ordinate after removing the points associated with the jump. That way $C_{ag}$ appears to tend to unity, as $t/t_w\rightarrow 1$, in a continuous fashion. Such a transformation does not alter the outcomes of the analysis that we perform below. It, in fact, brings better visual clarity over the relevant range. Nice collapse of data for different $t_w$ values is observed. It appears from these plots that we are reasonably away from the finite-size affected region \cite{Midya2014,Midya2015,Das2017}. Note that in the finite-size affected situation data for different $t_w$ will deviate from the master curve. The solid line in this figure represents a power-law decay with the exponent value mentioned in the figure. For large values of $t/t_w$, the simulation data are reasonably consistent with this line. However, to derive more accurate information on the decay of $C_{ag}$, below we carry out certain advanced analysis.

\begin{figure} [h]
\includegraphics*[width=0.48\textwidth]{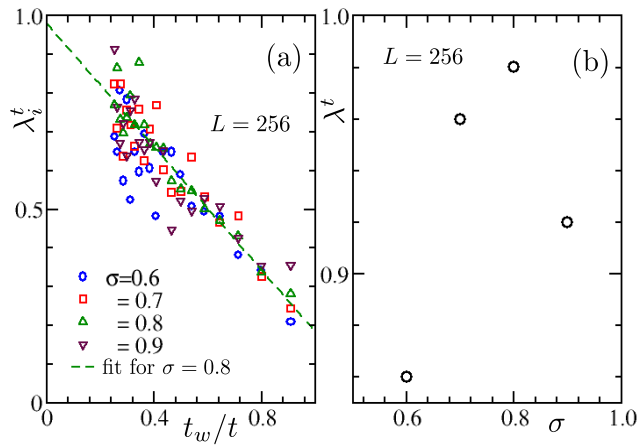}
\caption{ (a) The instantaneous exponent, $\lambda^t_i$, is shown versus $t_w/t$. Results from four $\sigma$ values have been displayed. The broken line is a linear fit to the data set corresponding to $\sigma=0.8$. This provides $\lambda^t=0.98$. (b) Plot of $\lambda^t$, obtained from the fits discussed in (a), versus $\sigma$.
}
\label{lambda_diff_sig}
\end{figure}

\begin{figure} [b]
\includegraphics*[width=0.48\textwidth]{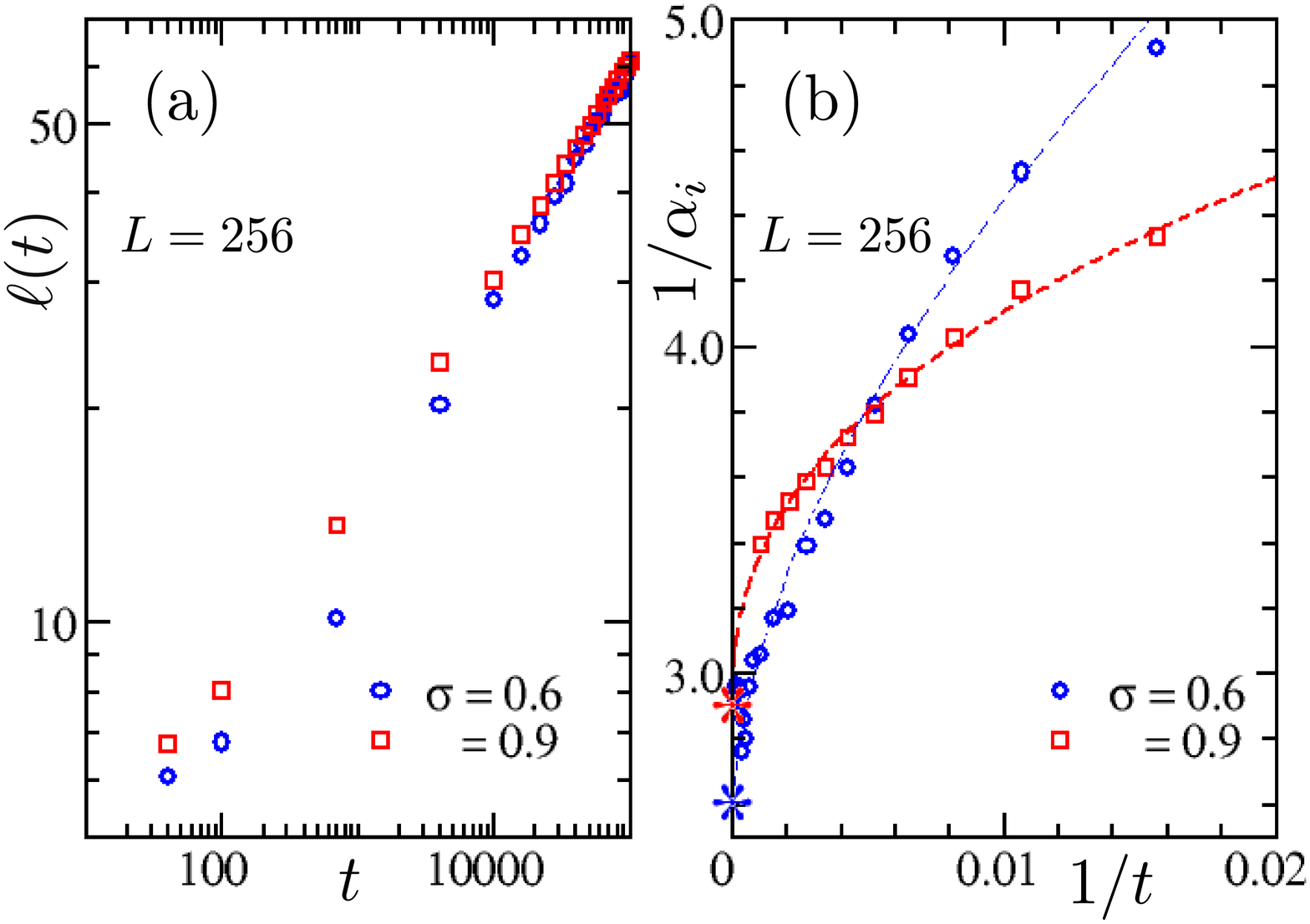}
\caption{$(a)$ Plots of average domain lengths with the variation of time. Results for two values of $\sigma$, viz., $0.6$ and $0.9$, are included. These are shown in a double-log scale. $(b)$ The inverse of the instantaneous exponents $\alpha_i$, for the same sets of data as in $(a)$, are presented versus $1/t$. The broken lines are guides to the eyes, showing possible convergences of the data sets to the  theoretical values $2+\sigma$ which are marked on the ordinate by asterisk ({\makebox{\rotatebox{30}{\large $\ast$}}}). The data at very late times have been discarded for better visualization of the convergences.}
\label{domain}
\end{figure}

We calculate the instantaneous exponent \cite{Fisher1988,Midya2014,Midya2015,Huse1986,Amar1988} $\lambda^t_i$ as
\begin{equation}
    \lambda_i^t=-\dfrac{d\ln C_{ag}(t,t_w)}{d\ln (t/t_w)},
\label{lambda_i}
\end{equation}
by anticipating that in the large $t/t_w$ limit $C_{ag}$ indeed decays in a power-law fashion. 
In Fig. \ref{lambda_diff_sig}(a) we present $\lambda_i^t$ as a function of $t_w/t$. Here we have included results from several values of $\sigma$. These show linear trend. There exists no discernable quantitative differences among the plots for different $\sigma$. The broken line there is a fit of the simulation data for $\sigma=0.8$ to the linear form $\lambda^t_i=\lambda^t+at_w/t$, $a$ being a constant. This exercise provides $\lambda^t=0.98$. In part (b) of Fig. \ref{lambda_diff_sig} we show thus obtained values of $\lambda^t$, versus $\sigma$. No particular dependence is evident. The variations appear random, being consistent with the above statement concerning the absence of any noticeable difference.

\begin{figure} [h]
\includegraphics*[width=0.48\textwidth]{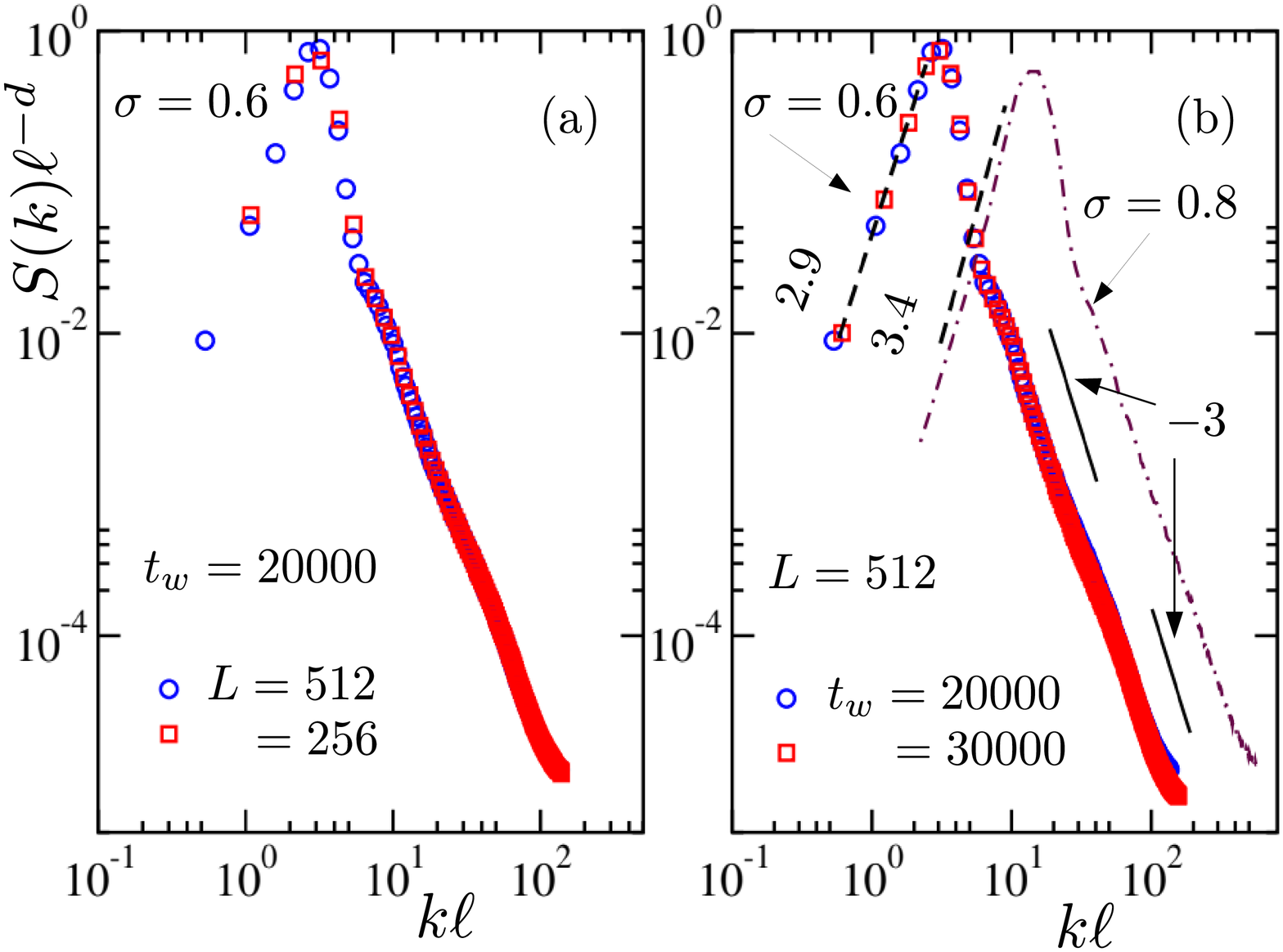}
\caption{
(a) Scaled structure factor, $S(k,t)\ell^{-d}$, at $t_w=20000$, is plotted versus $k\ell$, $k$ being the wave number, for systems of two different sizes, viz., $L=256$ and $512$, with $\sigma=0.6$. (b) Same scaling plots as in (a) are shown (with symbols) for two different $t_w$ values with $L=512$. The dotted-dashed line is the corresponding plot for $\sigma=0.8$. For better clarity of the features, the plot for $\sigma=0.8$ is shifted. The solid lines are proportional to $k^{-3}$. The dashed lines represent power-laws with exponents $\beta=2.9$ and $3.4$.   
}
\label{strft_fig}
\end{figure}

Thus, we treat $\lambda^t$ as practically independent of $\sigma$ in this range and obtain its value by averaging over the numbers for different $\sigma$. This way we estimate $\lambda^t=0.93$. To obtain $\lambda$, from $\lambda^t$, we need $\alpha$. While Eq. (\ref{alpha_con}) provides the $\sigma$-dependence of $\alpha$, which was confirmed by the recent simulations of M\"{u}ller et al. \cite{Muller2022}, here we revisit this issue. Before moving to that we metion that $\lambda$ may have $\sigma $-dependence in this range. But the random appearances of the values indicate that the dependence is weak.


In Fig. \ref{domain}(a) we show plots of $\ell$ versus $t$, on a double-log scale, for two values of $\sigma$, viz. $0.6$ and $0.9$. The growth, after early transients, appears stronger for the smaller value of $\sigma$. In part (b) of Fig. \ref{domain}, we show the instantaneous exponent for $\ell$, viz. $\alpha_i$ ($=d \ln \ell/d \ln t$), with the variation of $t$. Difference between the two cases is clearly visible. These outcomes, implying higher $\alpha$ for smaller $\sigma$, are in agreement with those in Ref. \cite{Muller2022}. Convergences of $\alpha_i$ to the theoretical values of $\alpha$ can also be appreciated.
\begin{figure} [t]
\includegraphics*[width=0.48\textwidth]{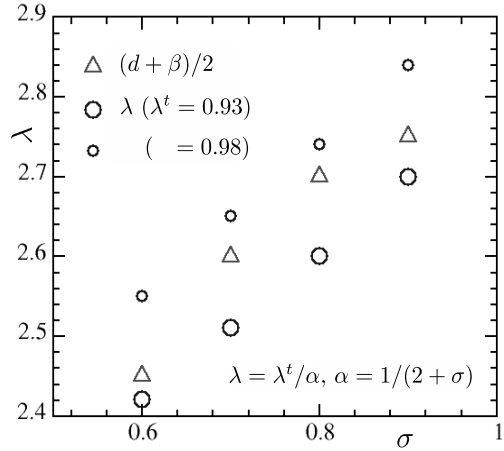}
\caption{The bound in Eq. (\ref{yrd_eq}) is plotted versus $
\sigma$ (see the triangles). The circles represent our estimated values of $\lambda$, with the variation of $\sigma$. This quantity is calculated as $\lambda= \lambda^t(2+\sigma)$. In one case we have considered $\lambda^t=0.93$ and in another, $\lambda^t=0.98$. See text for details.}
\label{yrd_fig}
\end{figure}

In Fig.\ref{strft_fig}(a) we show scaling plots \cite{Bray2002,Marko1995} of the equal time structure factor, $S(k,t_w)$, for $t_w=20000$, using data from different system sizes, viz, $L=256$ and $512$. The chosen value of $\sigma$ is $0.6$. While the data for the two system sizes agree with each other, the set from the larger system appears more useful with respect to the identification of short wave number ($k$) behaviour. In Fig. \ref{strft_fig}(b), we show results from $L=512$. For $\sigma=0.6$ results for two different times are shown in symbols. Nice scaling collapse is observed, confirming the validity of the scaling form $S(k,t_w)=\ell^d \tilde{S}(k\ell)$, $\tilde{S}$ being a master function. The small $k$ behaviour appears to be of power-law \cite{Yeung1988} type: $\sim\ k^\beta$, with $\beta = 2.9$. This way have extracted $\beta$ for several values of $\sigma$. At large $k$, $S(k,t_w)$ is expected to obey the Porod law \cite{Bray2002} $k^{-3}$. Indeed, this appears to be the case. Interestingly, as opposed to the nearest-neighbour case \cite{Bray2002}, one more decay step appears that also is consistent with $k^{-3}$. This multiple-step behaviour goes away with the increase of $\sigma$. See the plot for $\sigma=0.8$ in Fig. \ref{strft_fig}(b) for which the above mentioned multiple-step decay is absent as in Ref. \cite{Muller2022}. 

Yeung, Rao and Desai \cite{Yeung1996} provided an expression for the lower bounds on the values of $\lambda$:
\begin{equation}
    \lambda \ge \dfrac{d+\beta}{2}.
    \label{yrd_eq}
\end{equation}
In Fig. \ref{yrd_fig} we have plotted these bounds as a function of $\sigma$. These are compared with the obtained values of $\lambda$. Recall that we have used the formula $\lambda=\lambda^t(2+\sigma)$. For $\lambda^t=0.93$, the average value, the data set is represented by the bigger circles. Even though it appears that there exists consistent violation, the differences with the lower bounds are within $2\%$. In fact, if we use $\lambda^t=0.98$, instead of $0.93$, the value of $\lambda^t$ for $\sigma=0.8$, the agreements of $\lambda$ with the lower bounds are almost perfect. 
\section{Summary}
In conclusion, we have studied the kinetics of phase separation \cite{Bray2002} in the long-range Ising model \cite{Bray1993}. We have presented results on aging phenomena obtained via Monte Carlo simulations \cite{Landau2005} in space dimension $2$, for several values of the interaction range parameter $\sigma$. It appears that with the increase of $\sigma$, the aging exponent $\lambda$ increases. 

We have compared the values of $\lambda$ with the lower bounds predicted in Ref. \cite{Yeung1996}. It seems that the bounds provide the values of $\lambda$ quite accurately. In a previous study \cite{Midya2015}, the value of $\lambda$ for the nearest-neighbour Ising model in $d=2$ was estimated to be approximately $3.6$. For rather high values of $\sigma$, we have checked that our estimates are consistent with this. However, we are uncertain whether such a crossover occurs at $\sigma=1$, which is the boundary for the crossover of the growth exponent, between short-range and long-range classes. 

We have also discussed results on domain growth. Our results on this quantity are consistent with the theoretical predictions \cite{Fisher1972} and very recently published simulation reports by M\"{u}ller et al. \cite{Muller2022}. 
\section{Acknowledgement}
We acknowledge computation times in the Param Yukti supercomputer, located in JNCASR, under National Supercomputing Mission.



\bibliographystyle{revtex}
\bibliography{library_lrim_aging}
\end{document}